\def\beq#1{\begin{equation}\label{#1}}
\def\eeq{\end{equation}}
\def\beqa#1{\begin{eqnarray}\label{#1}}
\def\eeqa{\end{eqnarray}}

\def\fun#1#2{\lower3.6pt\vbox{\baselineskip0pt\lineskip.11pt
        \ialign{$\mathsurround=0pt#1\hfill##\hfil$\crcr#2\crcr\sim\crcr}}}

\def\xi{{{\bf x}^b}}

\documentclass
[onecolumn,aps,nofootinbib,superscriptaddress,amsmath]{revtex4}
\usepackage{epsfig}

\newcommand{\be}{\begin{equation}}
\newcommand{\ee}{\end{equation}}
\newcommand{\ba}{\begin{eqnarray}}
\newcommand{\ea}{\end{eqnarray}}

\begin{document}
\input{epsf.sty}

\title{Effects of a time-varying color-luminosity parameter $\beta$ on the cosmological constraints of modified gravity models}

\author{Shuang Wang}
\email{wjysysgj@163.com}
\affiliation{Department of Physics, College of Sciences, Northeastern University, Shenyang 110004, China}

\author{Yong-Zhen Wang}
\email{w_avin@163.com}
\affiliation{Department of Physics, College of Sciences, Northeastern University, Shenyang 110004, China}

\author{Xin Zhang\footnote{Corresponding author}}
\email{zhangxin@mail.neu.edu.cn}
\affiliation{Department of Physics, College of Sciences, Northeastern University, Shenyang 110004, China}
\affiliation{Center for High Energy Physics, Peking University, Beijing 100080, China}

\begin{abstract}
It has been found that, for the Supernova Legacy Survey three-year (SNLS3) data,
there is strong evidence for the redshift-evolution of color-luminosity parameter $\beta$.
In previous studies, only dark energy (DE) models are used to explore the effects of a time-varying $\beta$ on parameter estimation.
In this paper, we extend the discussions to the case of modified gravity (MG),
by considering Dvali-Gabadadze-Porrati (DGP) model, power-law type $f(T)$ model and exponential type $f(T)$ model.
In addition to the SNLS3 data, we also use the latest Planck distance priors data,
the galaxy clustering (GC) data extracted from Sloan Digital Sky Survey (SDSS) data release 7 (DR7) and Baryon Oscillation Spectroscopic Survey (BOSS),
as well as the direct measurement of Hubble constant $H_0$ from the Hubble Space Telescope (HST) observation.
We find that, for both cases of using the supernova (SN) data alone and using the combination of all data, 
adding a parameter of $\beta$ can reduce $\chi^2$ by $\sim$ 36 for all the MG models,
showing that a constant $\beta$ is ruled out at 6$\sigma$ confidence level (CL).
Moreover, we find that a time-varying $\beta$
always yields a larger fractional matter density $\Omega_{m0}$ and a smaller reduced Hubble constant $h$;
in addition, it significantly changes the shapes of 1$\sigma$ and 2$\sigma$ confidence regions of various MG models,
and thus corrects systematic bias for the parameter estimation.
These conclusions are consistent with the results of DE models,
showing that $\beta$'s evolution is completely independent of the cosmological models in the background.
Therefore, our work highlights the importance of considering the evolution of $\beta$ in the cosmology-fits.

\end{abstract}

\maketitle

\section{Introduction}

In recent 16 years,
cosmic acceleration \cite{Riess98,spergel03,Tegmark04,Komatsu09,Percival10,Drinkwater10}
has become one of the most important issues in modern cosmology.
To explain this puzzle,
one can either introduce an unknown energy component
(i.e. dark energy (DE) \cite{quint,phantom,k,Chaplygin,ngcg,tachyonic,HDE,hessence,YMC,hscalar,cq,others1,others2,others3,others4}),
or modify Einstein's general relativity (i.e. modified gravity (MG) \cite{SH,PR,DGP,GB,Galileon,FR,FT1,FT2,FRT}).
For recent reviews, see \cite{CST,FTH,Linder,CK,Uzan,Tsujikawa,NO,LLWW,CFPS,YWBook}.

One of the most powerful probes to illuminate the mystery of cosmic acceleration
is Type Ia supernovae (SNe Ia).
Several high-quality supernova (SN) datasets had been released in recent years \cite{Union,Constitution,Union2,Union2.1}.
The Supernova Legacy Survey three-year (SNLS3) data \cite{Guy10} were released in 2010.
Soon after, using SNLS3 dataset,
Conley et al. \cite{Conley11} and Sullivan et al. \cite{Sullivan11}
presented the SN-only cosmological results and the joint cosmological constraints, respectively.
Unlike other SN group, the SNLS team treated two important quantities,
stretch-luminosity parameter $\alpha$ and color-luminosity parameter $\beta$ of SNe Ia,
as free model parameters.

A critical challenge is the control of the systematic uncertainties of SNe Ia.
One of the most important factors that yield systematic uncertainties is the potential SN evolution,
i.e. the possibility for the redshift evolution of $\alpha$ and $\beta$.
So far, it is found that $\alpha$ is still consistent with a constant,
but the hints for the evolution of $\beta$ have been found in \cite{Astier06,Kessler09,Marriner11,Scolnic1,Scolnic2}.
In \cite{Mohlabeng}, using a linear $\beta(z) = \beta_0 + \beta_1 z$,
Mohlabeng and Ralston studied the case of Union2.1 dataset
and found that $\beta$ deviates from a constant at 7$\sigma$ confidence levels (CL).
In \cite{WangWang}, Wang \& Wang found that, for the SNLS3 data,
$\beta$ increases significantly with $z$ at the 6$\sigma$ CL;
moreover, they proved that this conclusion is insensitive to the lightcurve fitter models,
or the functional form of $\beta(z)$ assumed.
These studies show that the evolution of $\beta$ is a common phenomenon for various SN datasets,
and should be taken into account seriously.

It is very important to study the effects of a time-varying $\beta$ on parameter estimation of cosmological models.
In \cite{WangNew}, Wang, Li \& Zhang explored this issue
by considering the $\Lambda$-cold-dark-matter ($\Lambda$CDM) model, the $w$CDM model, and the Chevallier-Polarski-Linder (CPL) model.
Then, in \cite{WangNew2}, Wang, Geng, Hu \& Zhang studied the case of holographic dark energy (HDE) model,
which is a physically plausible DE candidate based on the holographic principle.
Next, in \cite{WangNew3}, Wang, Wang, Geng \& Zhang extended the discussion
to the case of considering the interaction between dark sectors.
It is found that, for all these models, $\beta$ deviates from a constant at $\sim$ 6$\sigma$ CL;
in addition, a time-varying $\beta$ will significantly change the confidence ranges of various cosmological parameters.

It must be stressed that, in previous studies, only DE models are adopted to explore the issue of varying $\beta$.
To do a comprehensive analysis on the cosmological consequences of a time-varying $\beta$,
it is necessary to extend the discussions to the case of MG,
which is another important approach to explaining cosmic acceleration.
So in this paper, we explore the effects of a time-varying $\beta$ on the cosmological constraints of three popular MG models,
including Dvali-Gabadadze-Porrati (DGP) model \cite{DGP} and two $f(T)$ models \cite{FT1,FT2}.
In addition to the SNLS3 data,
we also use the Planck distance prior data \cite{WangWangCMB} of the cosmic microwave background (CMB),
the galaxy clustering (GC) data
from Sloan Digital Sky Survey (SDSS) data release 7 (DR7) \cite{ChuangWang12} and Baryon Oscillation Spectroscopic Survey (BOSS) \cite{Chuang13},
as well as the direct measurement of Hubble constant $H_0$ from the Hubble Space Telescope (HST) observation \cite{Riess11}.

We describe our method in Sec.~II, present our results in Sec.~III, and conclude in Sec.~IV.
In this paper, we assume today's scale factor $a_{0}=1$, thus the redshift $z=a^{-1}-1$.
The subscript ``0'' always indicates the present value of the corresponding quantity,
and the natural units are used.

\section{Methodology}
\label{sec:method}

In this section, we introduce the theoretical models we considered
and the observational data we used in this paper.

\subsection{Theoretical models}

In this paper, we consider a Friedmann-Lematre-Robertson-Walker Universe with a non-zero spatial curvature.
We investigate three popular MG models, including the DGP model, the power-law type $f(T)$ model, and the exponential type $f(T)$ model.

\begin{itemize}
 \item DGP model
\end{itemize}

The DGP model is a braneworld model \cite{DGP},
where gravity is altered at immense distances by slow leakage of gravity off from our three-dimensional universe.
For this model, the dimensionless Hubble parameter $E(z)\equiv H(z)/H_{0}$ is given by
\begin{eqnarray}
E(z) = \left[ \sqrt{\Omega_{m0}(1+z)^3+\Omega_{r0}(1+z)^4+\Omega_{k0}(1+z)^2+\Omega_{rc}}+\sqrt{\Omega_{rc}}\right],
\end{eqnarray}
where $\Omega_{rc} = (1-\Omega_{m0}-\Omega_{r0}-\Omega_{k0})^2/4$.
Here $\Omega_{m0}$, $\Omega_{r0}$ and $\Omega_{k0}$
are the present fractional densities of matter, radiation and curvature, respectively.
In addition, we have $\Omega_{r0}=\Omega_{m0} / (1+z_{\rm eq})$,
and $z_{\rm eq}=2.5\times 10^4 \Omega_{m0} h^2 (T_{\rm cmb}/2.7\,{\rm K})^{-4}$ with 
$T_{\rm cmb}=2.7255\,{\rm K}$.

%%%%%%%%%%%%%%%%%%%%%%%%%%%%%%%%%%%%%%%%%%%%%%%%%%%%%%%%%%%%%%%%%%%%%%%%%%%%%%%%%%%%%%%%%%%%%%%%%%%%%%%%%%%%%%%%%%%%%%%%%%%%%%%%%%%%%%%%%%%%%%

\begin{itemize}
 \item $f(T)$ models
\end{itemize}

In the $f(T)$ gravity theory \cite{FT1,FT2},
the torsion scalar $T$ in the Lagrangian density is replaced by a generalized function $T + f(T)$,
then the corresponding action $S$ can be written as
\begin{equation}\label{fT:action}
S=\frac{1}{16\pi G}\int d^{4}x\sqrt{-g}[T+f(T)]+S_{m}+S_{r}+S_{k},
\end{equation}
where $S_{m}$, $S_{r}$ and $S_{k}$ are the actions of matter, radiation and curvature, respectively.
Since the torsion scalar $T$ and the Hubble expansion rate $H$ satisfy the relation $T=-6H^2$,
the modified Friedmann equation can be written as \cite{FT2}
\begin{equation}\label{fT:F.e.}
H^{2}=\frac{8\pi G}{3}(\rho_{m}+\rho_{r}+\rho_{k})-\frac{f}{6}-2H^{2}f_{T}.
\end{equation}
Here $\rho_{m}$, $\rho_{r}$ and $\rho_{k}$ denote the energy densities of matter, radiation and curvature, respectively;
beside, $f_{T}\equiv df/dT$ is the derivative of $f(T)$ with respect to $T$.
Making use of the Hubble constant $H_0$, Eq. (\ref{fT:F.e.}) can be rewritten as
\begin{equation}\label{fT:E}
E(z)^{2}=\frac{1}{1+2f_{T}}\big(\Omega_{m0}(1+z)^{3}+\Omega_{r0}(1+z)^{4}+\Omega_{k0}(1+z)^{2}-\frac{f}{6H_{0}^{2}}\big).
\end{equation}

Here we consider two $f(T)$ models:
one is a power law form model proposed in \cite{FT1}, the other is an exponential form model proposed by Linder \cite{FT2}.
For simplicity, hereafter we will call them $f(T)_{PL}$ model and $f(T)_{EXP}$ model, respectively.

The $f(T)_{PL}$ model assumes the following ansatz of $f(T)$,
\begin{equation} \label{eq:fT1}
f(T) = \alpha (-T)^{n}.
\end{equation}
Here $n$ is a free model parameter.
Using Eq. (\ref{fT:F.e.}), the value of $\alpha$ can be fixed as
\begin{equation}
\alpha = (6 H_{0}^{2})^{1-n} \frac{1-\Omega_{m0}-\Omega_{r0}-\Omega_{k0}}{2n-1}.
\end{equation}
For the $f(T)_{PL}$ model, Eq. (\ref{fT:E}) becomes
\begin{equation}\label{fTpl:E}
E(z)^{2}=\Omega_{m0}(1+z)^{3}+\Omega_{r0}(1+z)^{4}+\Omega_{k0}(1+z)^{2}+(1-\Omega_{m0}-\Omega_{r0}-\Omega_{k0})E(z)^{2n}.
\end{equation}
Moreover, this equation can be rewritten as
\begin{equation}\label{fTpl:dEdz}
\frac{dE}{dz}=\frac{3\Omega_{m0}(1+z)^{2}+4\Omega_{r0}(1+z)^{3}+2\Omega_{k0}(1+z)}{2E-2n(1-\Omega_{m0}-\Omega_{r0}-\Omega_{k0})E^{2n-1}}.
\end{equation}
Making use of the initial condition $E(0) = 1$ and numerically solving Eq. (\ref{fTpl:dEdz}),
the evolution of $E(z)$ for the $f(T)_{PL}$ model can be easily obtained.

The $f(T)_{EXP}$ model adopts the following ansatz of $f(T)$ \cite{FT2},
\begin{equation} \label{fT2}
f(T) = m T_0 \big(1-e^{-p\sqrt{T/T_{0}}}\big).
\end{equation}
Here $p$ is a free model parameter, $T_0 = -6H^{2}_{0}$, and
\begin{equation} \label{c}
m=\frac{1-\Omega_{m0}-\Omega_{r0}-\Omega_{k0}}{1-(1+p)e^{-p}}.
\end{equation}
For the $f(T)_{EXP}$ model, Eq. (\ref{fT:E}) becomes
\begin{equation}\label{fTexp:E}
E(z)^{2}=\Omega_{m0}(1+z)^{3}+\Omega_{r0}(1+z)^{4}+\Omega_{k0}(1+z)^{2}+m\Big(1-\big(1+pE(z)\big)e^{-pE(z)}\Big).
\end{equation}
Moreover, this equation can be rewritten as
\begin{equation}\label{fTexp:dEdz}
\frac{dE}{dz}=\frac{3\Omega_{m0}(1+z)^{2}+4\Omega_{r0}(1+z)^{3}+2\Omega_{k0}(1+z)}{2E-mp^{2}E(z)e^{-pE(z)}}.
\end{equation}
Making use of the initial condition $E(0) = 1$ and numerically solving Eq. (\ref{fTexp:dEdz}),
the evolution of $E(z)$ for the $f(T)_{EXP}$ model can be obtained, too.

\subsection{Observational data}

In this subsection, we introduce how to calculate the $\chi^2$ function for SNLS3 data in detail.

For the SNLS3 sample, the observable is $m_B$, which is the rest-frame peak B-band magnitude of the SN.
By considering three functional forms (linear case, quadratic case, and step function case),
Wang \& Wang \cite{WangWang} showed that the evolutions of $\alpha$ and $\beta$
are insensitive to functional form of $\alpha$ and $\beta$ assumed.
So in this paper,
we just adopt a constant $\alpha$ and a linear $\beta(z) = \beta_{0} + \beta_{1} z$.
Then, the predicted magnitude of a SN becomes
\be
m_{\rm mod}=5 \log_{10}{\cal D}_L(z)
- \alpha (s-1) +\beta(z) {\cal C} + {\cal M},
\ee
where $s$ and ${\cal C}$ are the stretch measure and the color measure for the SN light curve.
Here ${\cal M}$ is a parameter representing some combination of SN absolute magnitude $M$ and Hubble constant $H_0$.
It must be emphasized that,
to include host-galaxy information in the cosmological fits,
Conley et al. \cite{Conley11} split the SNLS3 sample based on host-galaxy stellar mass at $10^{10} M_{\odot}$,
and made ${\cal M}$ to be different for the two samples.
Therefore, unlike other SN samples, there are two values of ${\cal M}$, ${\cal M}_1$ and ${\cal M}_2$, for the SNLS3 data
(for the details, see the subsections $3.2$ and $5.8$ of \cite{Conley11}).
Moreover, Conley et al. removed ${\cal M}_1$ and ${\cal M}_2$ from cosmological fits by analytically marginalizing over them
(for more details, see the appendix C of \cite{Conley11},
as well as the the public code which is available at https://tspace.library.utoronto.ca/handle/1807/24512).
In this paper, we just follow the recipe of Ref. \cite{Conley11};
following Ref. \cite{Conley11}, we do not report the values of ${\cal M}_1$ and ${\cal M}_2$.

The luminosity distance ${\cal D}_L(z)$ is defined as
\be
{\cal D}_L(z)\equiv H_0 (1+z_{\rm hel}) r(z),
\ee
where $z$ and $z_{\rm hel}$ are the CMB restframe and heliocentric redshifts of SN.
In addition, the comoving distance $r(z)$ is given by
\be
\label{eq:r(z)}
 r(z)=H_0^{-1}\, |\Omega_{k0}|^{-1/2} {\rm sinn}\big (|\Omega_{k0}|^{1/2}\, \Gamma(z)\big ),
\ee
where $\Gamma(z)=\int_0^z\frac{dz'}{E(z')}$,
and ${\rm sinn}(x)=\sin(x)$, $x$, $\sinh(x)$ for $\Omega_{k0}<0$, $\Omega_{k0}=0$, and $\Omega_{k0}>0$ respectively.

For a set of $N$ SNe with correlated errors, the $\chi^2$ function is
\be
\label{eq:chi2_SN}
\chi^2_{SN}=\Delta \mbox{\bf m}^T \cdot \mbox{\bf C}^{-1} \cdot \Delta\mbox{\bf m},
\ee
where $\Delta m \equiv m_B-m_{\rm mod}$ is a vector with $N$ components,
and $\mbox{\bf C}$ is the $N\times N$ covariance matrix of the SN, given by
\be
\mbox{\bf C}=\mbox{\bf D}_{\rm stat}+\mbox{\bf C}_{\rm stat}+\mbox{\bf C}_{\rm sys}.
\ee
$\mbox{\bf D}_{\rm stat}$ is the diagonal part of the statistical uncertainty, given by \cite{Conley11}
\ba
\mbox{\bf D}_{{\rm stat},ii}&=&\sigma^2_{m_B,i}+\sigma^2_{\rm int}
+ \sigma^2_{\rm lensing}+ \sigma^2_{{\rm host}\,{\rm correction}} + \left[\frac{5(1+z_i)}{z_i(1+z_i/2)\ln 10}\right]^2 \sigma^2_{z,i} \nonumber\\
&& +\alpha^2 \sigma^2_{s,i}+\beta(z_i)^2 \sigma^2_{{\cal C},i} + 2 \alpha C_{m_B s,i} - 2 \beta(z_i) C_{m_B {\cal C},i} -2\alpha \beta(z_i) C_{s {\cal C},i},
\ea
where $C_{m_B s,i}$, $C_{m_B {\cal C},i}$, and $C_{s {\cal C},i}$
are the covariances between $m_B$, $s$, and ${\cal C}$ for the $i$-th SN,
$\beta_i=\beta(z_i)$ are the values of $\beta$ for the $i$-th SN.
Notice that $\sigma^2_{z,i}$ includes a peculiar velocity residual of 0.0005
(i.e., 150$\,$km/s) added in quadrature.
Following Ref. \cite{Conley11}, we fix the intrinsic scatter $\sigma_{int}$ to ensure that $\chi^2/dof=1$.
Varying $\sigma_{int}$ could have a significant impact on parameter estimation, see \cite{Kim2011} for details.

We define $\mbox{\bf V} \equiv \mbox{\bf C}_{\rm stat} + \mbox{\bf C}_{\rm sys}$,
where $\mbox{\bf C}_{\rm stat}$ and $\mbox{\bf C}_{\rm sys}$
are the statistical and systematic covariance matrices, respectively.
After treating $\beta$ as a function of $z$,
$\mbox{\bf V}$ is given in the form,
\ba
\mbox{\bf V}_{ij}&=&V_{0,ij}+\alpha^2 V_{a,ij} + \beta_i\beta_j V_{b,ij} +\alpha V_{0a,ij} +\alpha V_{0a,ji}
-\beta_j V_{0b,ij} -\beta_i V_{0b,ji} -\alpha \beta_j V_{ab,ij} - \alpha \beta_i V_{ab,ji}.
\ea
It must be stressed that, while $V_0$, $V_{a}$, $V_{b}$, and $V_{0a}$
are the same as the ``normal'' covariance matrices
given by the SNLS data archive, $V_{0b}$, and $V_{ab}$ are {\it not} the same as the ones given there.
This is because the original matrices of SNLS3 are produced by assuming $\beta$ is constant.
We have used the $V_{0b}$ and $V_{ab}$ matrices for the ``Combined'' set
that are applicable when varying $\beta(z)$ (A.~Conley, private communication, 2013).

To improve the cosmological constraints,
we also use some other cosmological observations,
including the Planck distance prior data \cite{WangWangCMB},
the galaxy clustering (GC) data extracted from SDSS DR7 \cite{ChuangWang12} and BOSS \cite{Chuang13},
as well as the direct measurement of Hubble constant $H_0=73.8\pm 2.4~{\rm km/s/Mpc}$ from the HST observations \cite{Riess11}.
For the details of including these data into the $\chi^2$ analysis, see Refs. \cite{WangNew,WangNew2,WangNew3}.
Now the total $\chi^2$ function is
\be
\chi^2=\chi^2_{SN}+\chi^2_{CMB}+\chi^2_{GC}+\chi^2_{H0}.
\ee

Finally, we perform an MCMC likelihood analysis \cite{COSMOMC}
to obtain ${\cal O}$($10^6$) samples for each model considered in this paper.

\section{Results}

\subsection{Evolution of $\beta$}
\label{sec:varbeta}

In this subsection,
we explore the evolution of $\beta$ by considering the DGP model, the $f(T)_{PL}$ model and the $f(T)_{EXP}$ model.
As mentioned above, to explore the evolution of $\beta$,
we study the case of constant $\alpha$ and linear $\beta(z) = \beta_{0} + \beta_{1} z$;
for comparison, the case of constant $\alpha$ and constant $\beta$ is also taken into account.

\begin{itemize}
 \item SN-only case
\end{itemize}

Firstly, we discuss the results given by the SN data alone.
Notice that the reduced Hubble constant $h$ has been marginalized during the $\chi^2$ fitting process of SNe Ia.

In Table \ref{table1}, we list the fitting results for various constant $\beta$ and linear $\beta(z)$ cases,
where only the SNLS3 SN data are used.
The most obvious feature of this table is that varying $\beta$ can significantly improve the fitting results of various MG models:
for all the models, adding a parameter of $\beta$ can reduce the best-fit values of $\chi^2$ by $\sim$ 36.
Based on the Wilk's theorem, 36 units of $\chi^2$ is equivalent to a Gaussian fluctuation of 6$\sigma$.
Therefore, for the case of using SNLS3 data alone,
a constant $\beta$ is ruled out at 6$\sigma$ CL for all the MG models.
This result is consistent with the results of dark energy cases \cite{WangNew,WangNew2,WangNew3},
showing that the evolution of $\beta$ is completely independent of the cosmological model in the background.
Therefore, by taking into account the MG models,
we further confirm the redshift-evolution of $\beta$ for the SNLS3 data.

\begin{table*}\tiny
\caption{Fitting results for various constant $\beta$ and linear $\beta(z)$ cases,
where only the SN data are used.}
\label{table1}
\begin{tabular}{ccccccccc}
\hline\hline &\multicolumn{2}{c}{DGP}&&\multicolumn{2}{c}{$f(T)_{PL}$}&&\multicolumn{2}{c}{$f(T)_{EXP}$} \\
           \cline{2-3}\cline{5-6}\cline{8-9}
Parameter  & Const $\beta$ & Linear $\beta(z)$ && Const $\beta$ & Linear $\beta(z)$ && Const $\beta$ & Linear $\beta(z)$ \\ \hline
$\alpha$           & $1.425^{+0.112}_{-0.100}$
                   & $1.409^{+0.108}_{-0.092}$&
                   & $1.424^{+0.107}_{-0.095}$
                   & $1.416^{+0.100}_{-0.098}$&
                   & $1.463^{+0.055}_{-0.150}$
                   & $1.392^{+0.098}_{-0.049}$\\

$\beta_0$          & $3.264^{+0.107}_{-0.110}$
                   & $1.450^{+0.345}_{-0.406}$&
                   & $3.259^{+0.142}_{-0.103}$
                   & $1.437^{+0.354}_{-0.387}$&
                   & $3.289^{+0.082}_{-0.122}$
                   & $1.466^{+0.235}_{-0.319}$\\

$\beta_1$          &
                   & $5.075^{+1.151}_{-0.923}$&
                   &
                   & $5.095^{+1.081}_{-0.938}$&
                   &
                   & $5.056^{+0.905}_{-0.636}$\\

$\Omega_{m0}$      & $0.121^{+0.075}_{-0.082}$
                   & $0.021^{+0.157}_{-0.011}$&
                   & $0.259^{+0.239}_{-0.249}$
                   & $0.052^{+0.324}_{-0.042}$&
                   & $0.468^{+0.050}_{-0.458}$
                   & $0.523^{+0.052}_{-0.513}$\\

$\Omega_{k0}$      & $0.0621^{+0.2375}_{-0.1947}$
                   & $0.3754^{+0.1361}_{-0.3647}$&
                   & $-0.7040^{+1.4111}_{-0.3698}$
                   & $0.5037^{+0.3332}_{-1.4052}$&
                   & $-1.0433^{+1.4011}_{-0.0751}$
                   & $-1.1034^{+1.7594}_{-0.1069}$\\

$n$                &
                   & &
                   & $1.2279^{+0.5949}_{-4.2276}$
                   & $-0.0334^{+1.3375}_{-2.9645}$&
                   &
                   & \\

$p$                &
                   & &
                   &
                   & &
                   & $ -0.5068^{+10.5054}_{-1.5496}$
                   & $0.0374^{+9.9595}_{-0.6623}$\\

\hline $\chi^{2}_{min}$  & 419.758  & 383.621 && 419.567 & 383.622 && 419.340 & 383.161  \\
\hline
\end{tabular}
\end{table*}

\begin{figure}
\includegraphics[scale=0.25, angle=0]{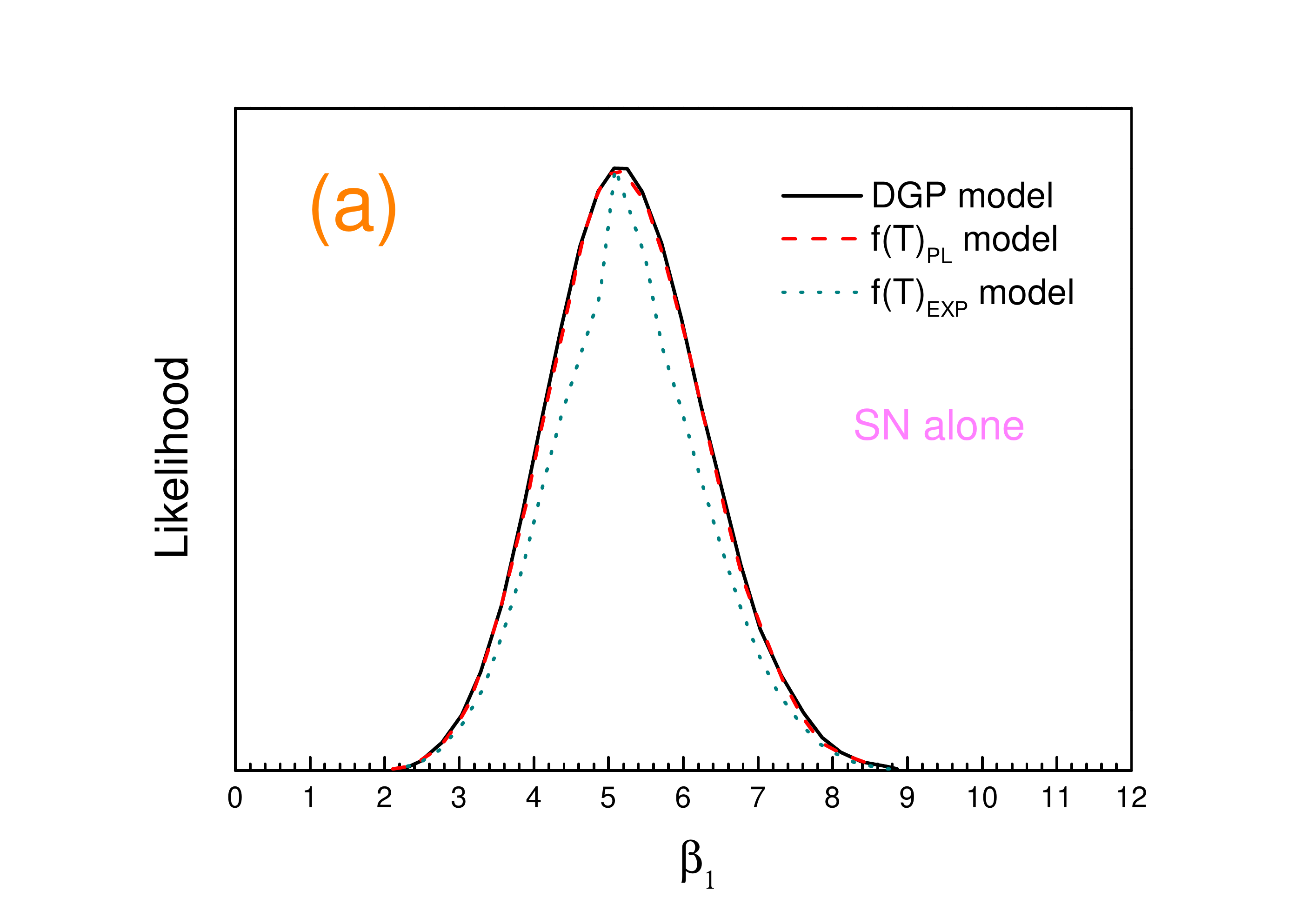}
\includegraphics[scale=0.25, angle=0]{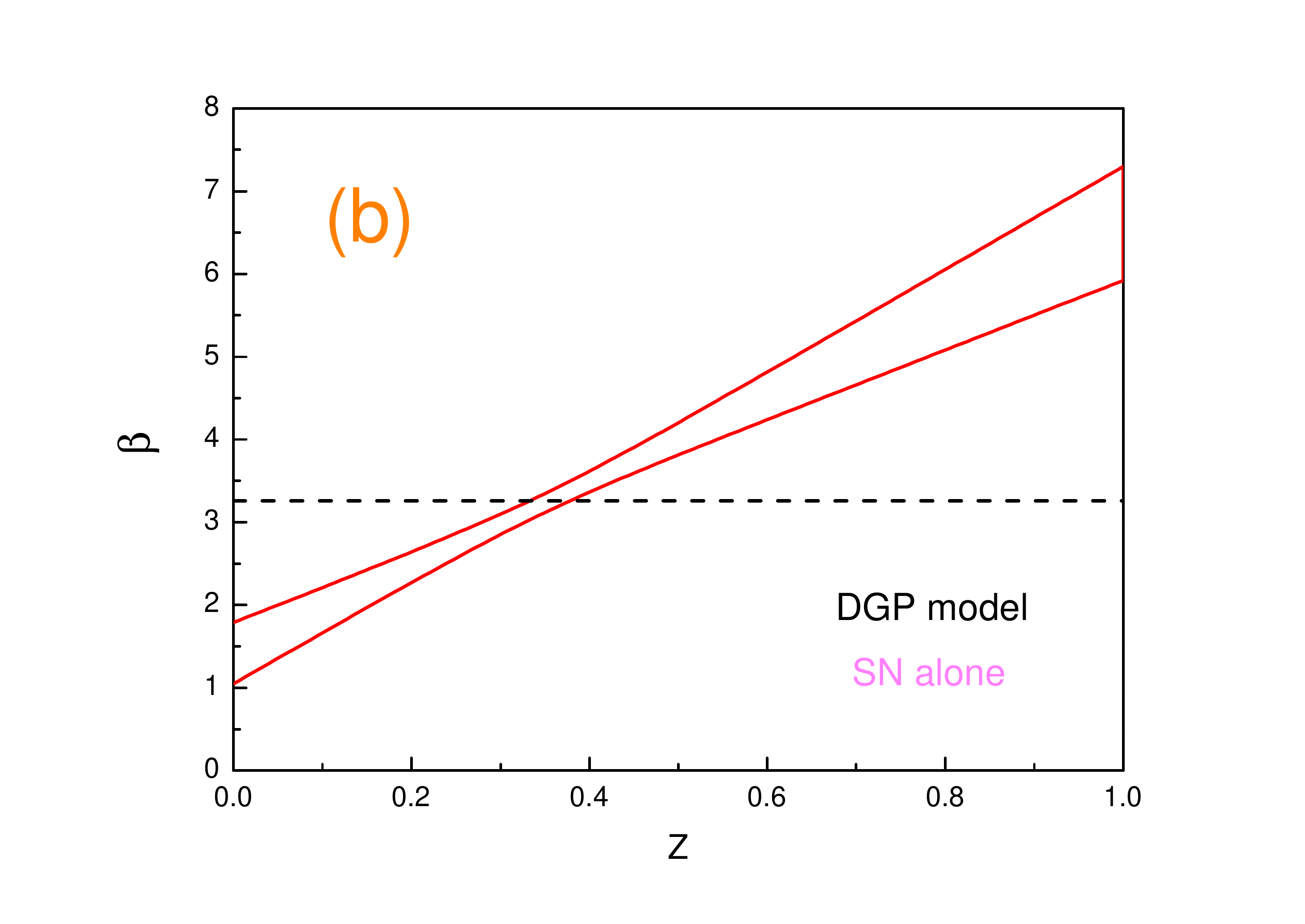}
\includegraphics[scale=0.25, angle=0]{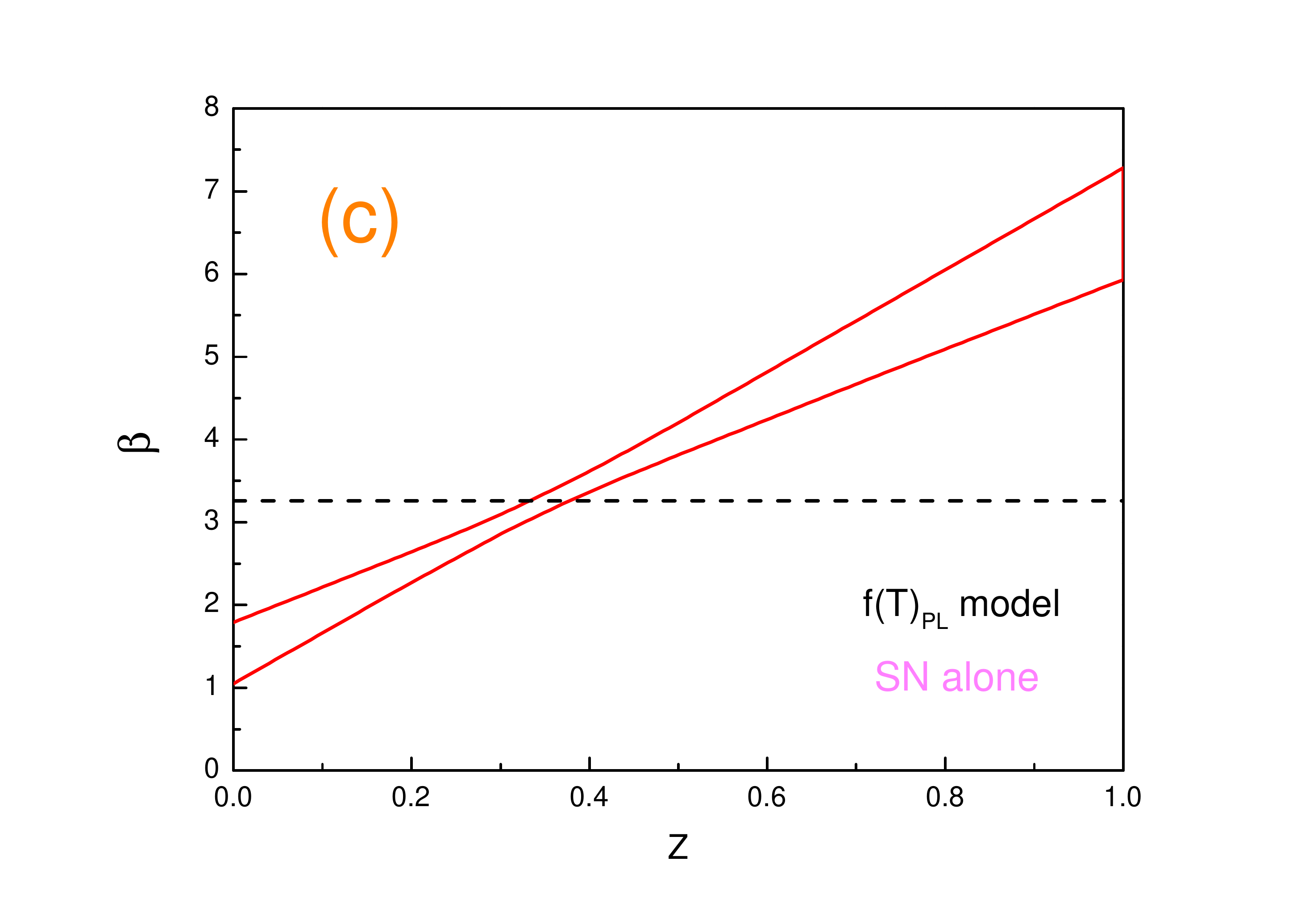}
\includegraphics[scale=0.25, angle=0]{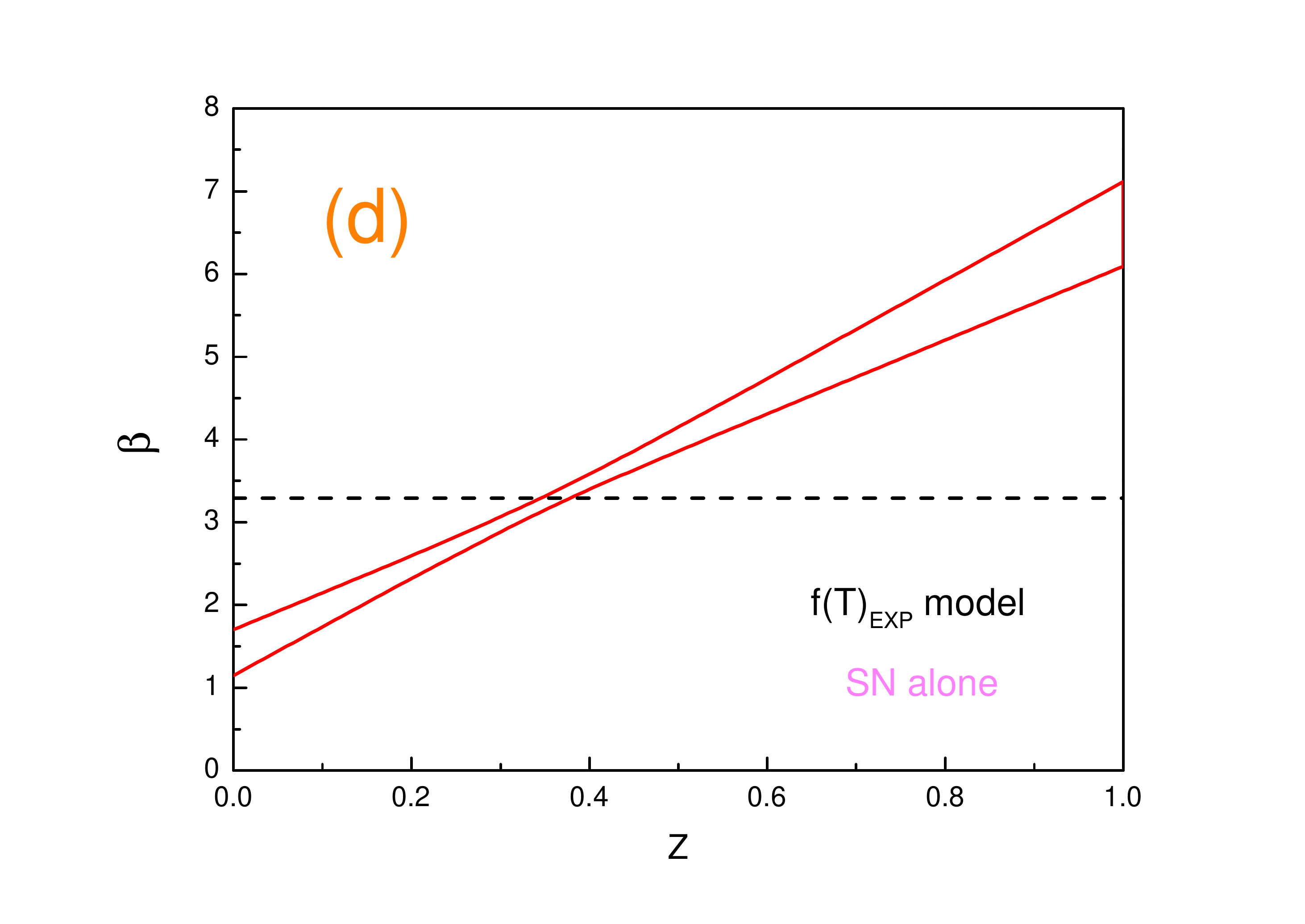}
\caption{\label{fig1}\footnotesize%
1D marginalized probability distributions of $\beta_{1}$ (panel a),
as well as 1$\sigma$ confidence constraints of $\beta(z)$ for the DGP (panel b), the $f(T)_{PL}$ (panel c), and the $f(T)_{EXP}$ model (panel d).
Only the SN data are used in the analysis. To make a comparison, the best-fit results of various constant $\beta$ cases are also plotted.}
\end{figure}

In Fig. \ref{fig1}, using the SN data alone,
we plot 1D marginalized probability distributions of $\beta_{1}$ (panel a),
as well as 1$\sigma$ confidence constraints of $\beta(z)$
for the DGP model (panel b), the $f(T)_{PL}$ model (panel c), and the $f(T)_{EXP}$ model (panel d).
To make a comparison, the best-fit results of various constant $\beta$ cases are also shown.
The panel A shows that, for all the MG models, $\beta_{1}>0$ at 6$\sigma$ CL;
while the panels B, C, and D show that, $\beta(z)$ rapidly increases with $z$.
These results further confirm that the evolution of $\beta$ is independent of the MG models,
showing that the importance of considering evolution of $\beta$ in the cosmology-fits.

It should be pointed out that the evolutionary behavior of $\beta(z)$ depends on the SN samples used.
In \cite{Mohlabeng}, Mohlabeng and Ralston found that, for the Union2.1 SN data, $\beta(z)$ decreases with $z$.
This is similar to the case of Pan-STARRS1 SN data \cite{Scolnic2}.
It is of great interest to study why different SN data give different evolutionary behaviors of $\beta(z)$,
and some numerical simulation studies may be required to solve this problem.
We will study this issue in future works.

\begin{itemize}
 \item SN+CMB+GC+$H_{0}$ case
\end{itemize}

Next, let us discuss the results given by the SN+CMB+GC+$H_{0}$ data.
It should be mentioned that, in order to use the Planck distance priors data,
two new model parameters, reduced Hubble parameter $h$ and radiation parameter $\Omega_{b}h^{2}$ must be added.

In Table \ref{table2}, we make a comparison for the fitting results of constant $\beta$ and linear $\beta(z)$ cases,
where the SN+CMB+GC+$H_{0}$ data are used.
Again, we see that varying $\beta$ can significantly improve the fitting results of various MG models:
for the $f(T)_{PL}$ and the $f(T)_{EXP}$ models,
adding a parameter of $\beta$ will reduce the best-fit values of $\chi^2$ by $\sim$ 36;
for the DGP model, adding a parameter of $\beta$ will reduce the best-fit value of $\chi^2$ by $\sim$ 47.
Therefore, the conclusion of $\beta_1\neq0$ still holds true for the SN+CMB+GC+$H_{0}$ case.
In addition, only using the SN data,
the $\chi^{2}_{min}$ value of DGP model is almost the same to the results of $f(T)_{PL}$ and $f(T)_{EXP}$ models;
once taking into account other observational data,
the $\chi^{2}_{min}$ value of DGP model becomes significantly larger than the results of $f(T)_{PL}$ and $f(T)_{EXP}$ models.
This implies that, for the DGP model,
the cosmological constraints given by the SN data is significantly inconsistent with the cosmological constraints given by other cosmological observations.
Therefore, we can conclude that the DGP model is strongly disfavored by the current cosmological observations;
this is consistent with the conclusions of many previous works \cite{Fang08,Rubin09,Li:2009jx,ZWS}.

\begin{table*}\tiny
\caption{Fitting results for various constant $\beta$ and linear $\beta(z)$ cases,
where the SN+CMB+GC+$H_0$ data are used.}
\label{table2}
\begin{tabular}{ccccccccc}
\hline\hline &\multicolumn{2}{c}{DGP}&&\multicolumn{2}{c}{$f(T)_{PL}$}&&\multicolumn{2}{c}{$f(T)_{EXP}$} \\
           \cline{2-3}\cline{5-6}\cline{8-9}
Parameter  & Const $\beta$ & Linear $\beta(z)$ && Const $\beta$ & Linear $\beta(z)$ && Const $\beta$ & Linear $\beta(z)$ \\ \hline
$\alpha$           & $1.406^{+0.112}_{-0.097}$
                   & $1.398^{+0.098}_{-0.087}$&
                   & $1.414^{+0.110}_{-0.086}$
                   & $1.394^{+0.120}_{-0.073}$&
                   & $1.433^{+0.089}_{-0.111}$
                   & $1.415^{+0.098}_{-0.096}$\\

$\beta_0$          & $3.250^{+0.123}_{-0.089}$
                   & $1.101^{+0.416}_{-0.310}$&
                   & $3.266^{+0.098}_{-0.105}$
                   & $1.506^{+0.309}_{-0.408}$&
                   & $3.243^{+0.116}_{-0.098}$
                   & $1.448^{+0.343}_{-0.341}$\\

$\beta_1$          &
                   & $6.063^{+0.887}_{-1.148}$&
                   &
                   & $4.950^{+1.045}_{-0.885}$&
                   &
                   & $5.093^{+1.006}_{-0.919}$\\

$\Omega_{m0}$      & $0.299^{+0.012}_{-0.011}$
                   & $0.303^{+0.012}_{-0.010}$&
                   & $0.264^{+0.013}_{-0.009}$
                   & $0.281^{+0.017}_{-0.011}$&
                   & $0.276^{+0.009}_{-0.011}$
                   & $0.279^{+0.010}_{-0.010}$\\

$\Omega_{k0}$      & $0.0310^{+0.0034}_{-0.0040}$
                   & $0.0303^{+0.0037}_{-0.0038}$&
                   & $-0.002^{+0.0044}_{-0.0044}$
                   & $0.0016^{+0.0050}_{-0.0051}$&
                   & $0.0050^{+0.0031}_{-0.0033}$
                   & $0.0051^{+0.0028}_{-0.0035}$\\

$n$                &
                   & &
                   & $-0.465^{+0.271}_{-0.300}$
                   & $-0.083^{+0.202}_{-0.236}$&
                   &
                   & \\

$p$                &
                   & &
                   &
                   & &
                   & $9.0888^{+0.9118}_{-3.8316}$
                   & $7.1324^{+2.8669}_{-2.7097}$\\

$h$                & $0.683^{+0.011}_{-0.012}$
                   & $0.678^{+0.010}_{-0.012}$&
                   & $0.731^{+0.011}_{-0.017}$
                   & $0.707^{+0.014}_{-0.021}$&
                   & $0.712^{+0.013}_{-0.011}$
                   & $0.710^{+0.012}_{-0.012}$\\

$\Omega_{b}h^{2}$  & $0.0224^{+0.00031}_{-0.00027}$
                   & $0.0224^{+0.00026}_{-0.00030}$&
                   & $0.0222^{+0.00032}_{-0.00025}$
                   & $0.0223^{+0.00028}_{-0.00027}$&
                   & $0.0223^{+0.00033}_{-0.00024}$
                   & $0.0223^{+0.00030}_{-0.00027}$\\

\hline $\chi^{2}_{min}$  & 455.965  & 408.834 && 421.941 & 386.965 && 425.410 & 388.878  \\
\hline
\end{tabular}
\end{table*}

\begin{figure}
\includegraphics[scale=0.25, angle=0]{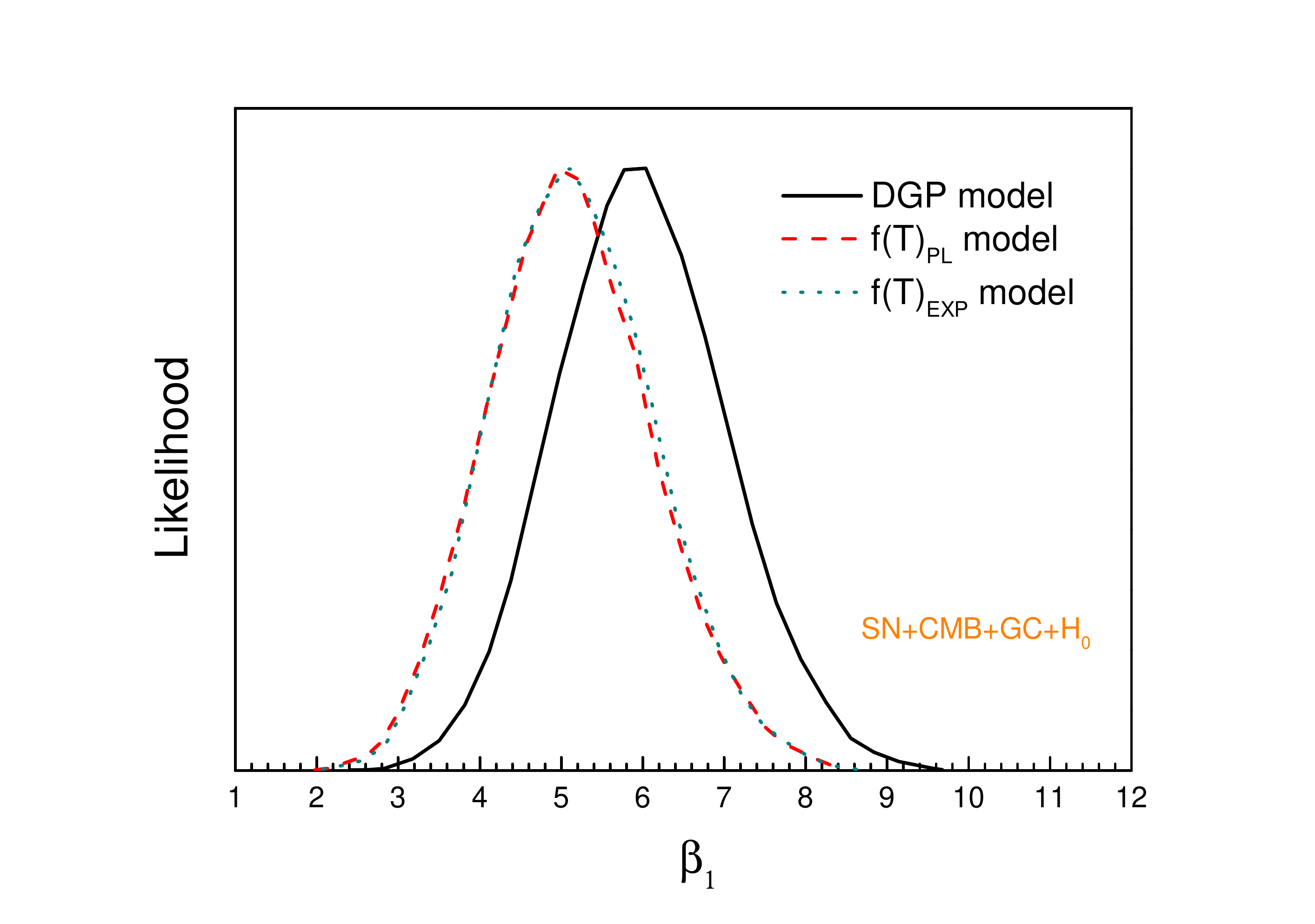}
\includegraphics[scale=0.25, angle=0]{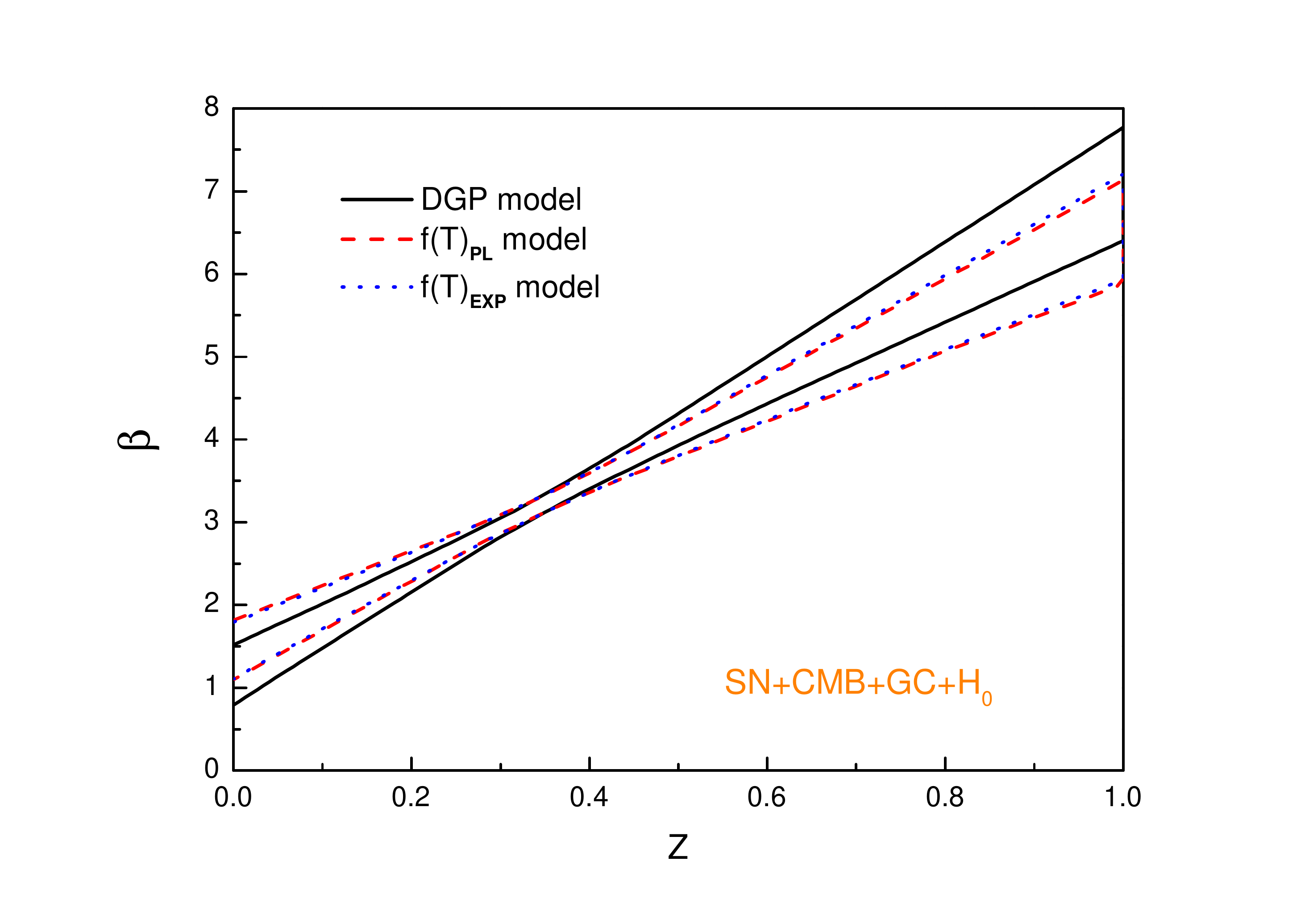}
\caption{\label{fig2}\footnotesize%
1D marginalized probability distributions of $\beta_{1}$ (left panel)
and 1$\sigma$ confidence constraints of $\beta(z)$ (right panel) for various MG models.
The SN+CMB+GC+$H_0$ data are used in the analysis.}
\end{figure}

In Fig. \ref{fig2}, using SN+CMB+GC+$H_0$ data,
we plot 1D marginalized probability distributions of $\beta_{1}$ (left panel)
and 1$\sigma$ confidence constraints of $\beta(z)$ (right panel) for various MG models.
The left panel shows that, for all the MG models, $\beta_{1}>0$ at 6$\sigma$ CL;
while the right panel show that $\beta(z)$ rapidly increases with $z$ for all the models.
This result is just the same to the result of SN-only case.
In addition, it is also consistent with the results of dark energy cases \cite{WangNew,WangNew2,WangNew3}.
Notice that the evolution of $\beta$ for the DGP model is slightly different from the results for the $f(T)_{PL}$ and $f(T)_{EXP}$ models;
this maybe due to the possible degeneracy between $\Omega_{m0}$ and $\beta(z)$.
But this slight difference has no influence on the conclusion of time-varying $\beta$.

\subsection{Effects of time-varying $\beta$}
\label{sec:betaeff}

In this subsection, we discuss the effects of varying $\beta$ on parameter estimations of various MG models.
For simplicity, in this subsection we just use the SN+CMB+GC+$H_0$ data.

\begin{itemize}
 \item DGP model
\end{itemize}

Firstly, let us discuss the results of DGP model.
An advantage of this model is that it has the same parameter number with the simplest $\Lambda$CDM model.

\begin{figure}
\includegraphics[scale=0.25, angle=0]{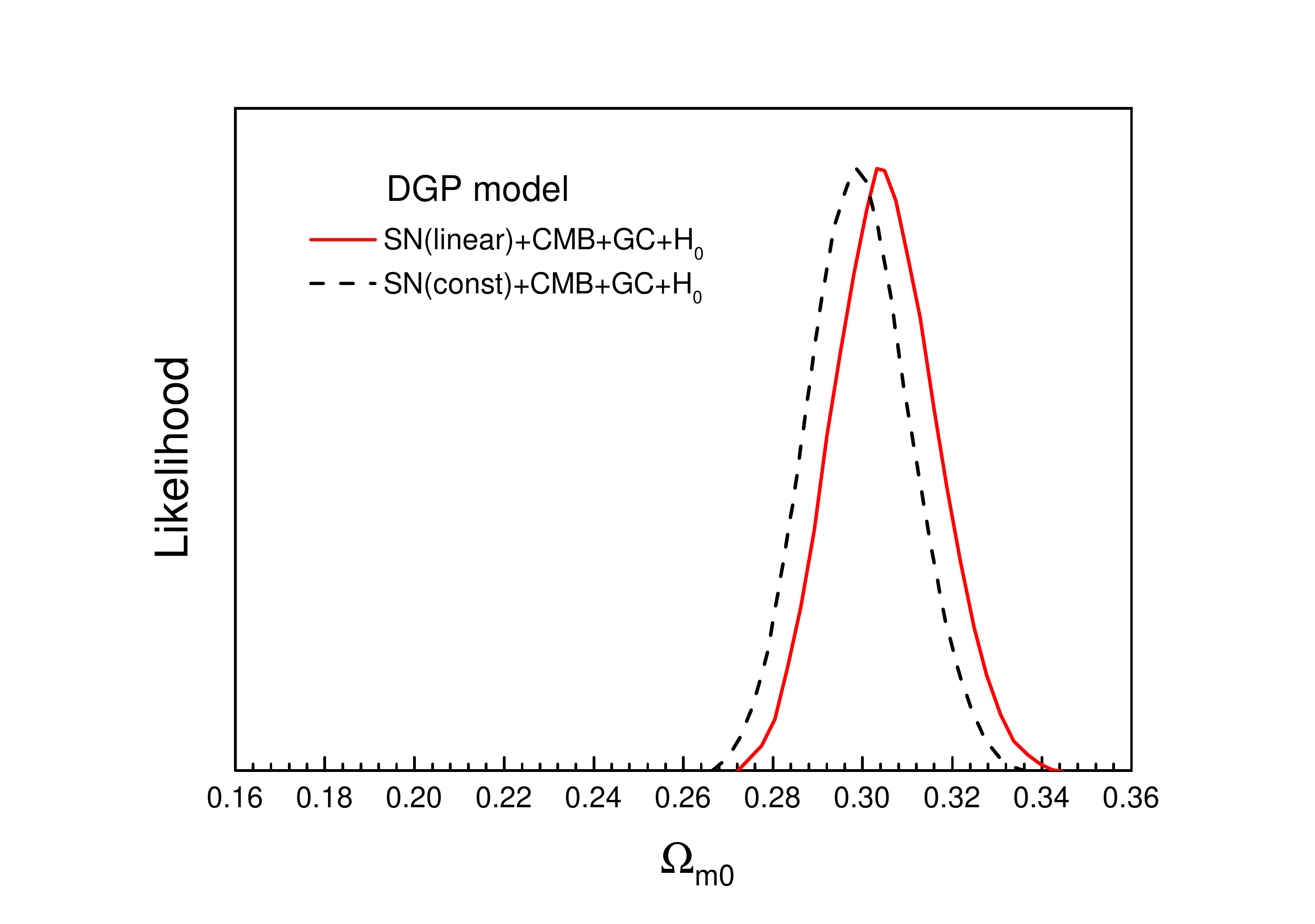}
\includegraphics[scale=0.25, angle=0]{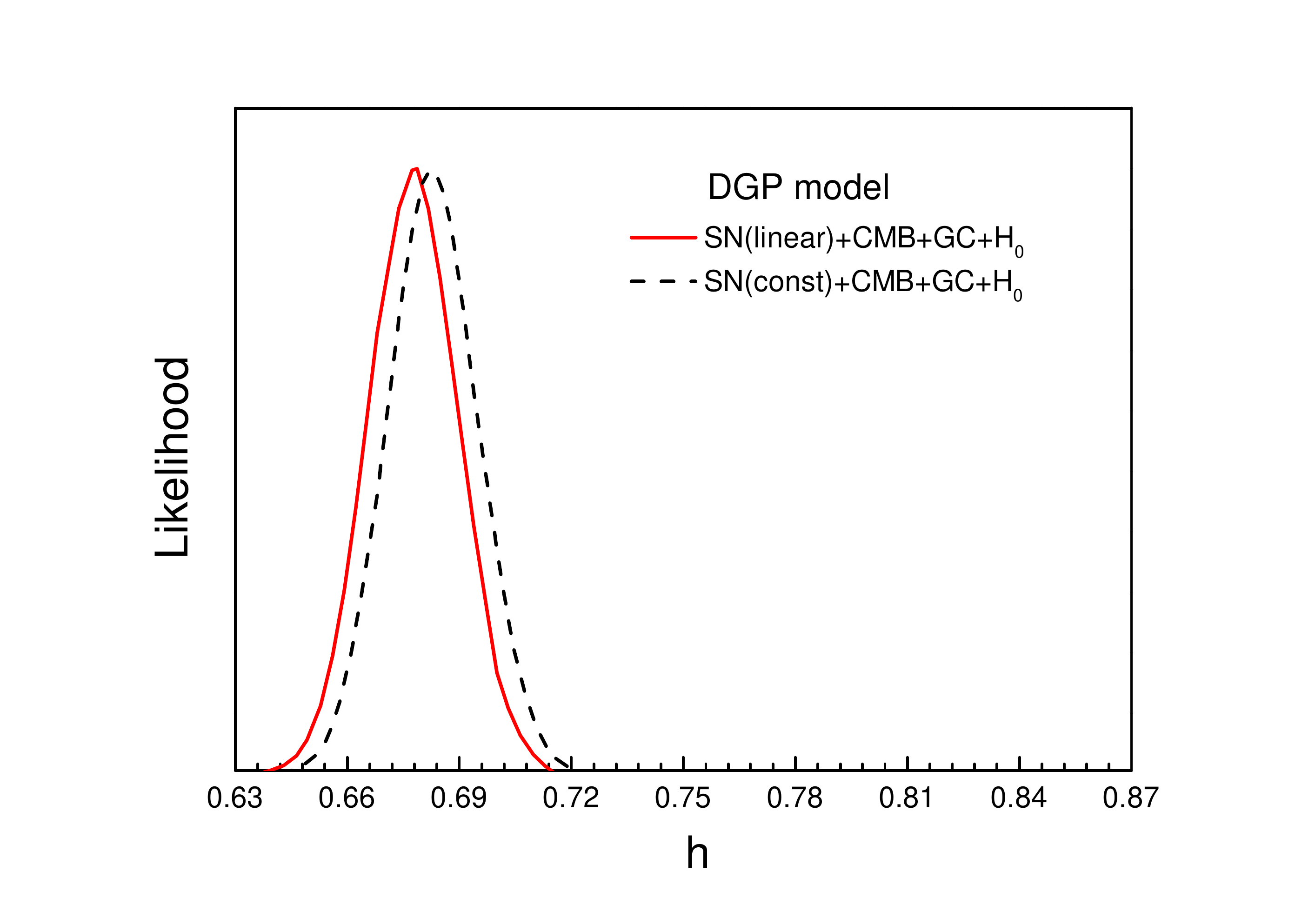}
\caption{\label{fig3}\footnotesize%
1D marginalized probability distributions of $\Omega_{m0}$ (left panel) and $h$ (right panel),
given by the SN+CMB+GC+$H_0$ data, for the DGP model.
Both the results of constant $\beta$ and linear $\beta(z)$ cases are presented.}
\end{figure}

In Fig. \ref{fig3},
we plot 1D marginalized probability distributions of $\Omega_{m0}$ (left panel) and $h$ (right panel), for the DGP model.
We find that varying $\beta$ yields a larger $\Omega_{m0}$ and a smaller $h$:
the best-fit results of constant $\beta$ case are $\Omega_{m0}=0.299$ and $h=0.683$,
while best-fit results of the linear $\beta(z)$ case are $\Omega_{m0}=0.303$ and $h=0.678$.
This result is consistent with the conclusions of dark energy cases \cite{WangNew,WangNew2,WangNew3}.

\begin{itemize}
 \item $f(T)_{PL}$ model
\end{itemize}

Then, let us turn to the case of $f(T)_{PL}$ model.
This model has an additional model parameter $n$.

\begin{figure}
\includegraphics[scale=0.25, angle=0]{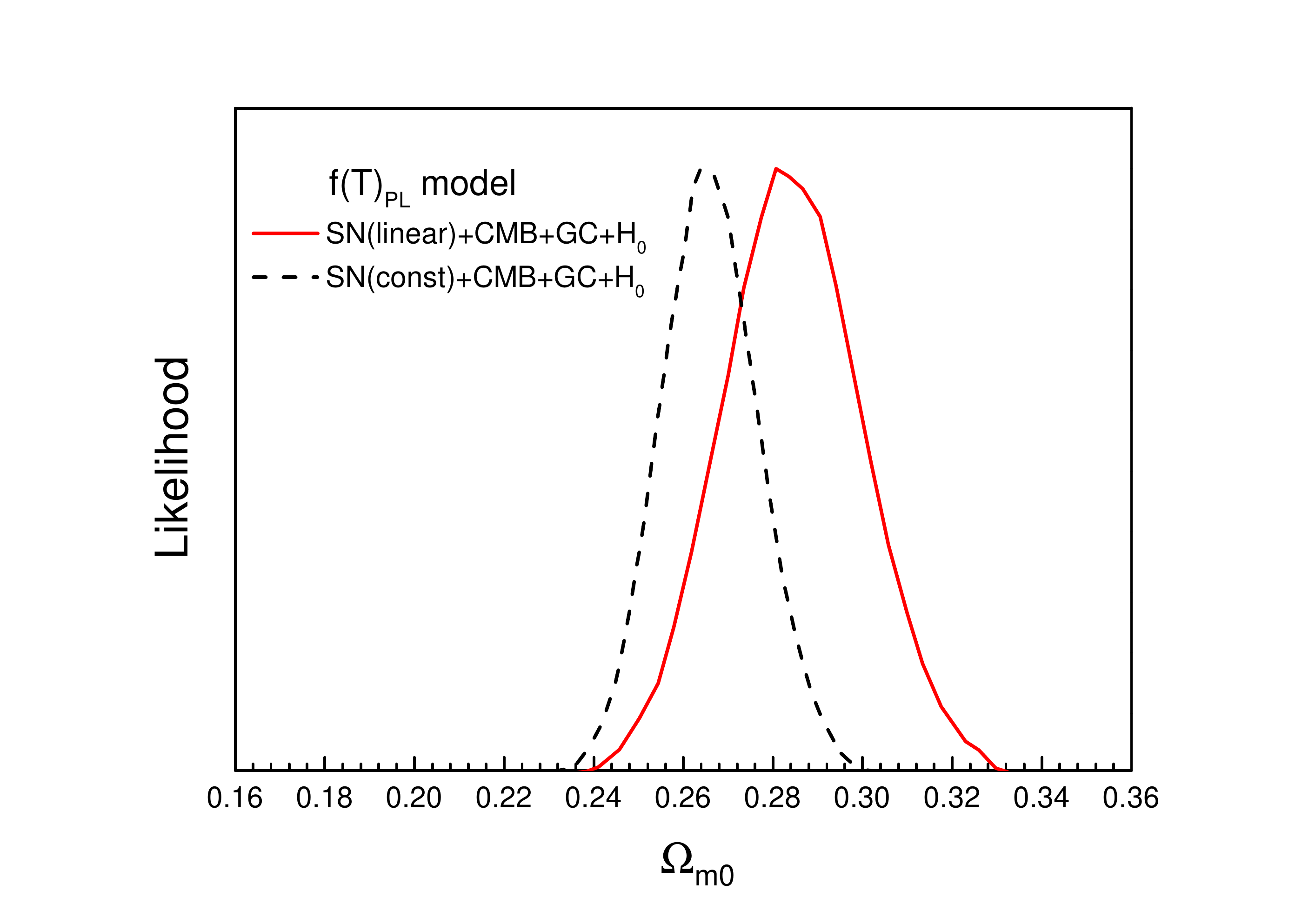}
\includegraphics[scale=0.25, angle=0]{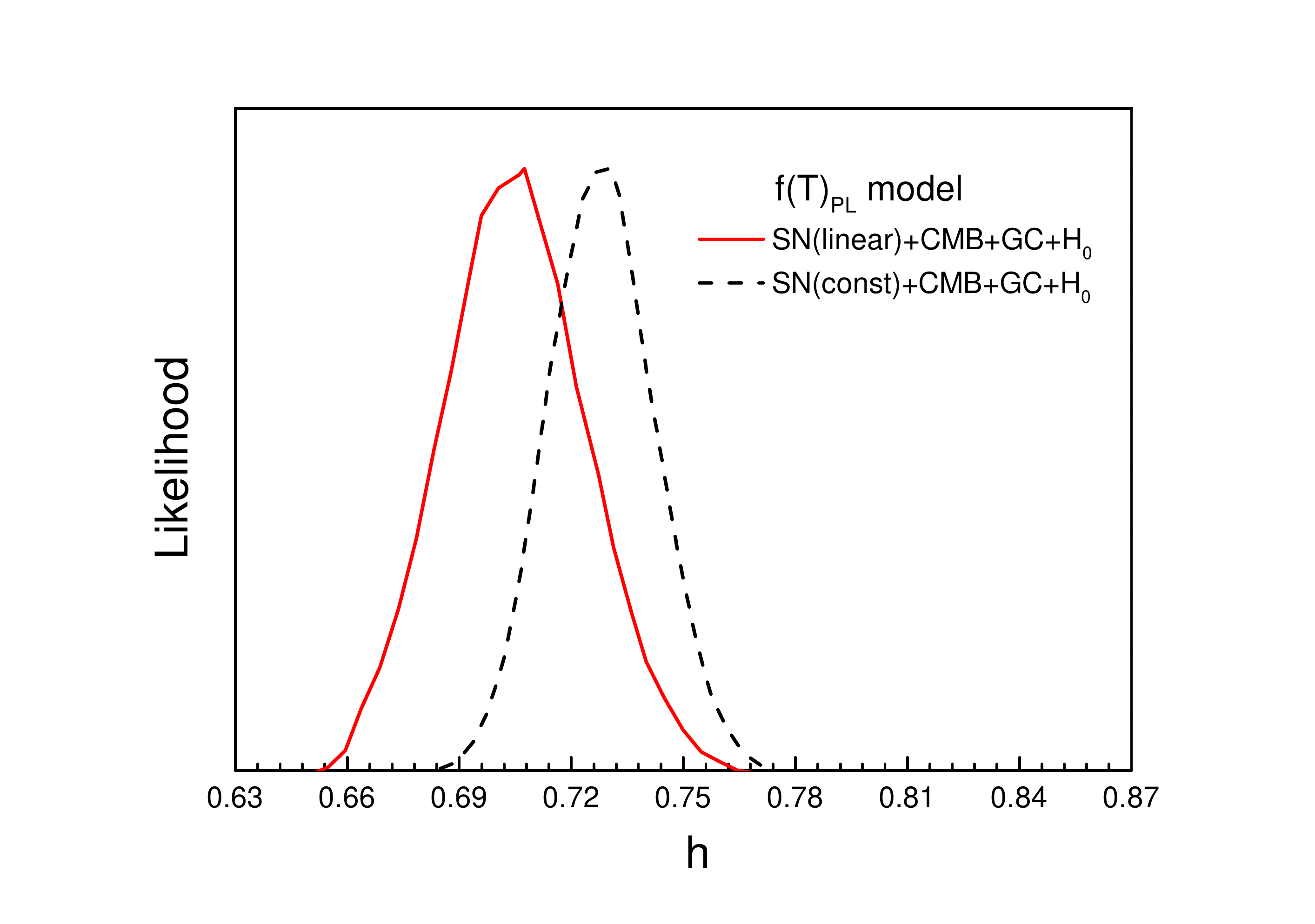}
\caption{\label{fig4}\footnotesize%
1D marginalized probability distributions of $\Omega_{m0}$ (left panel) and $h$ (right panel),
given by the SN+CMB+GC+$H_0$ data, for the $f(T)_{PL}$ model.
Both the results of constant $\beta$ and linear $\beta(z)$ cases are presented.}
\end{figure}

In Fig. \ref{fig4},
we plot 1D marginalized probability distributions of $\Omega_{m0}$ (left panel) and $h$ (right panel), for the $f(T)_{PL}$ model.
It can be seen that varying $\beta$ also yields a larger $\Omega_{m0}$ and a smaller $h$ for this case:
the best-fit results of constant $\beta$ case are $\Omega_{m0}=0.264$ and $h=0.731$,
while best-fit results of the linear $\beta(z)$ case are $\Omega_{m0}=0.281$ and $h=0.707$.
This result is consistent with the result of Fig. \ref{fig3}.

\begin{figure}
\psfig{file=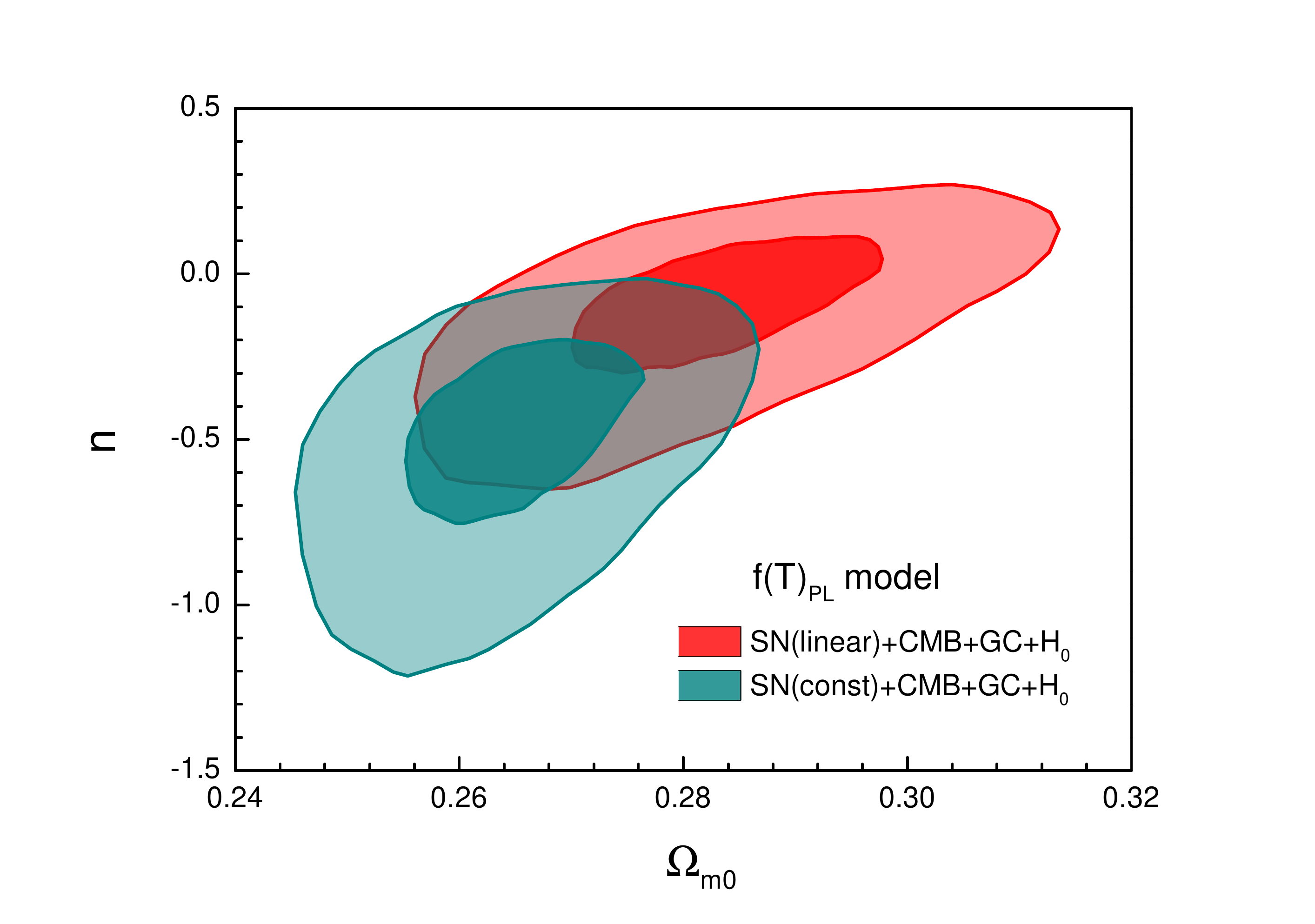,width=3.5in}\\
\caption{\label{fig5}\footnotesize%
The 1$\sigma$ and 2$\sigma$ confidence contours of $\{\Omega_{m0},n\}$, for the $f(T)_{PL}$ model.
Both the results of constant $\beta$ and linear $\beta(z)$ cases are presented.
}
\end{figure}

In Fig. \ref{fig5},
we plot the 1$\sigma$ and 2$\sigma$ confidence contours of $\{\Omega_{m0},n\}$, for the $f(T)_{PL}$ model.
From this figure, one can see that varying $\beta$ yields a larger $n$:
the best-fit value of constant $\beta$ case is $n=-0.465$,
while best-fit value of the linear $\beta(z)$ case is $n=-0.083$.
Moreover, it can be seen that a time-varying $\beta$ significantly change the shapes of 1$\sigma$ and 2$\sigma$ confidence regions;
this implies that ignoring the evolution of $\beta$ will cause systematic bias.

\begin{itemize}
 \item $f(T)_{EXP}$ model
\end{itemize}

Finally, we turn to the $f(T)_{EXP}$ model, which has an additional model parameter $p$.

\begin{figure}
\includegraphics[scale=0.25, angle=0]{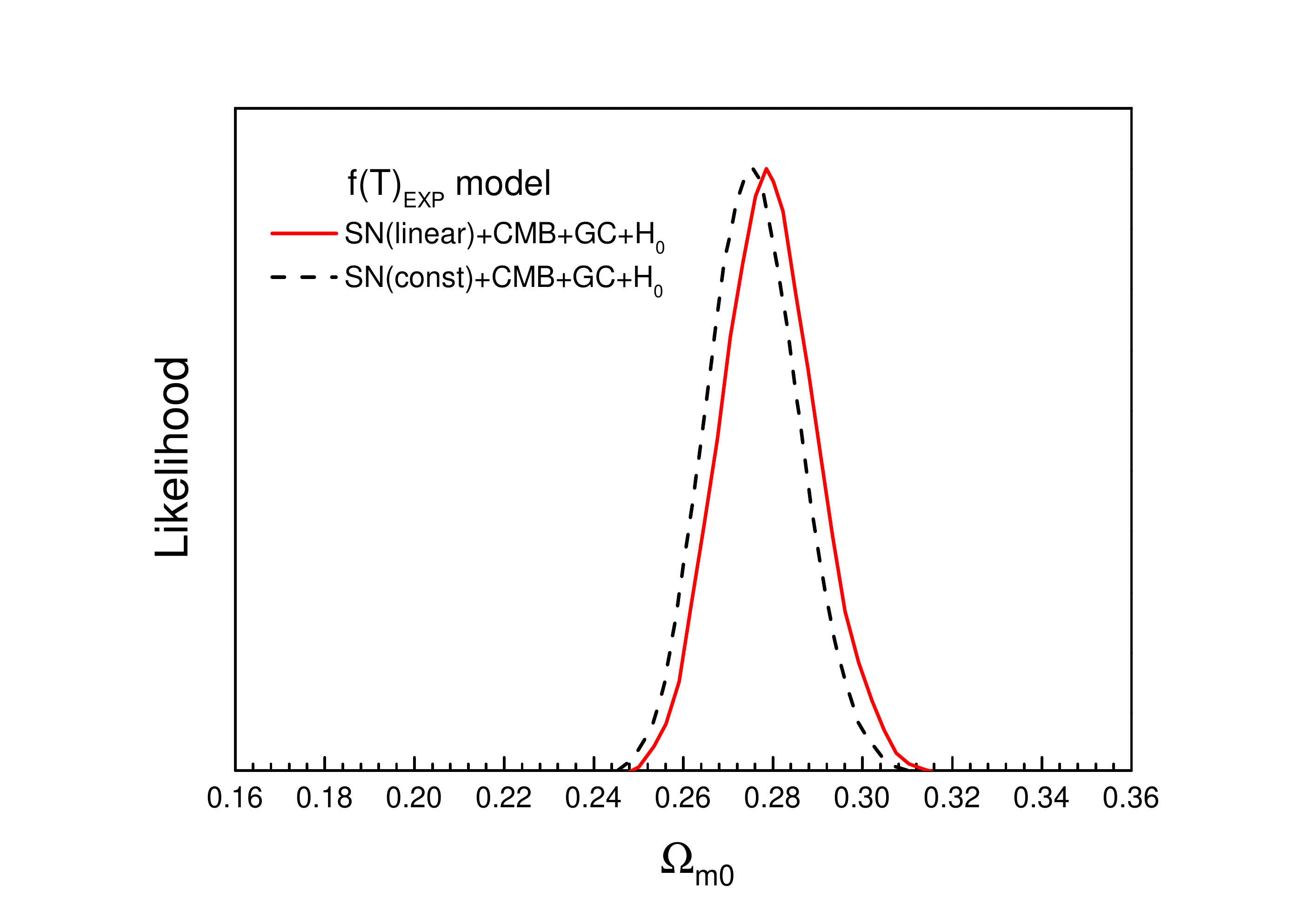}
\includegraphics[scale=0.25, angle=0]{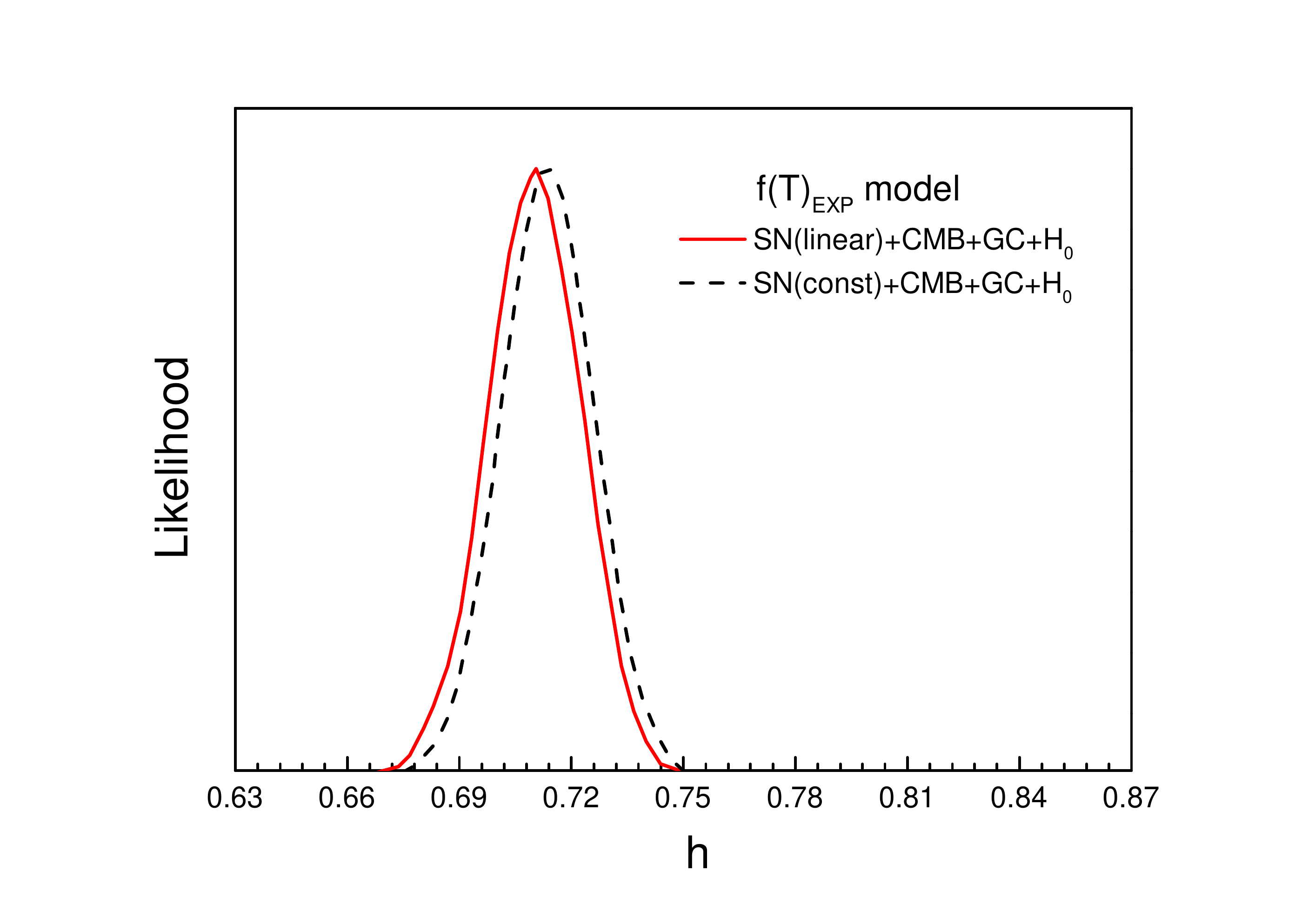}
\caption{\label{fig6}\footnotesize%
1D marginalized probability distributions of $\Omega_{m0}$ (left panel) and $h$ (right panel),
given by the SN+CMB+GC+$H_0$ data, for the $f(T)_{EXP}$ model.
Both the results of constant $\beta$ and linear $\beta(z)$ cases are presented.}
\end{figure}

In Fig. \ref{fig6},
we plot 1D marginalized probability distributions of $\Omega_{m0}$ (left panel) and $h$ (right panel),
for the $f(T)_{EXP}$ model.
Again, we see that varying $\beta$ yields a larger $\Omega_{m0}$ and a smaller $h$:
the best-fit results of constant $\beta$ case are $\Omega_{m0}=0.276$ and $h=0.712$,
while best-fit results of the linear $\beta(z)$ case are $\Omega_{m0}=0.279$ and $h=0.710$.
This result is consistent with the results of Fig. \ref{fig3} and Fig. \ref{fig4}.

\begin{figure}
\psfig{file=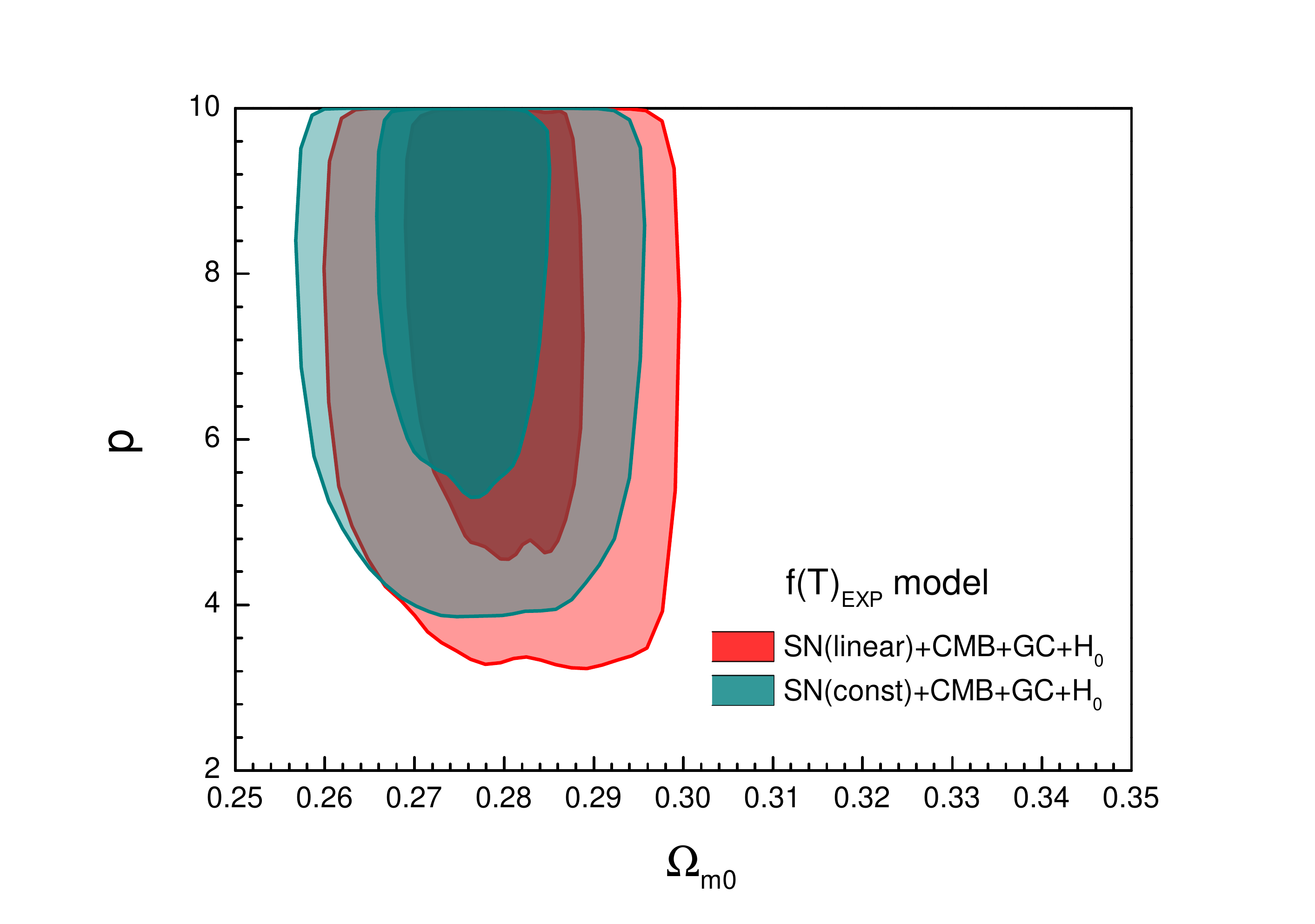,width=3.5in}\\
\caption{\label{fig7}\footnotesize%
The 1$\sigma$ and 2$\sigma$ confidence contours of $\{\Omega_{m0},p\}$, for the $f(T)_{EXP}$ model.
Both the results of constant $\beta$ and linear $\beta(z)$ cases are presented.
}
\end{figure}

In Fig. \ref{fig7},
we plot the 1$\sigma$ and 2$\sigma$ confidence contours of $\{\Omega_{m0},p\}$, for the $f(T)_{EXP}$ model.
It can be seen that varying $\beta$ yields a smaller $p$;
in addition, a time-varying $\beta$ will change the shapes of 1$\sigma$ and 2$\sigma$ confidence regions.

According to Figs. \ref{fig3},  \ref{fig4} and  \ref{fig6},
we can conclude that a time-varying $\beta$ always yields a larger $\Omega_{m0}$ and a smaller $h$.
In addition, based on Figs. \ref{fig5} and  \ref{fig7},
we can conclude that varying $\beta$ significantly changes the shapes of 1$\sigma$ and 2$\sigma$ confidence regions, and thus corrects systematic bias.
These two conclusions are independent of the cosmological models in the background.

\section{Discussion and summary}

In recent years,
the control of the systematic uncertainties of SNe Ia has drawn more and more attention.
One of the most important systematic uncertainties for SNe Ia is the potential SN evolution.
The hints for the evolution of $\beta$ have been found \cite{Astier06,Kessler09,Marriner11,Scolnic1,Scolnic2,Mohlabeng}.
In \cite{WangWang}, using the SNLS3 data,
Wang \& Wang found strong evidence for the redshift-evolution of $\beta$;
moreover, they proved that the evolution of $\beta$ is insensitive to the lightcurve fitter models,
or the functional form of $\beta(z)$ assumed.

It is clear that a time-varying $\beta$ will have significant impact on parameter estimation.
Adopting a constant $\alpha$ and a linear $\beta(z) = \beta_{0} + \beta_{1} z$,
Wang, Li \& Zhang \cite{WangNew} explored this issue by considering $\Lambda$CDM model, $w$CDM model, and CPL model.
Then, Wang, Geng, Hu \& Zhang \cite{WangNew2} studied this issue in the framework of HDE model,
which is a physically plausible DE candidate based on the holographic principle.
Soon after, Wang, Wang, Geng \& Zhang \cite{WangNew3} extended the corresponding discussion
to the case of considering the interaction between dark sectors.
It is found that, for all these models, $\beta$ deviates from a constant at $\sim$ 6$\sigma$ CL;
in addition, a time-varying $\beta$ will significantly change the confidence ranges of various cosmological parameters.

It must be stressed that, in previous studies, only DE models are adopted to explore the issue of varying $\beta$.
To do a comprehensive analysis on the cosmological consequences of a time-varying $\beta$,
it is necessary to extend the discussions to the case of MG.
So in this paper, we explore the effects of a time-varying $\beta$ on the cosmological constraints of three popular MG models,
including DGP model,  $f(T)_{PL}$ model and $f(T)_{EXP}$ model.
In addition to the SNLS3 SN data,
we also use the Planck distance priors data,
the GC data extracted from SDSS DR7 and BOSS,
as well as the direct measurement of Hubble constant $H_0$ from the HST observation.

In this paper,
we further confirm the evidence of redshift-evolution of $\beta$ for the SNLS3 data.
We find that, for both the cases of using the SN data alone and using the combination of all data,
adding a parameter of $\beta$ can reduce $\chi^2$ by $\sim$ 36 for all the MG models,
showing that a constant $\beta$ is ruled out at 6$\sigma$ CL.
Moreover, we find that a time-varying $\beta$ always yields a larger $\Omega_{m0}$ and a smaller $h$;
in addition, it significantly changes the shapes of 1$\sigma$ and 2$\sigma$ confidence regions of various MG models,
and thus corrects systematic bias for the parameter estimation.

The conclusions of our paper are consistent with the results of DE cases,
showing that the conclusion of time-varying $\beta$ holds true for both DE and MG models.
In other words, $\beta$'s evolution is completely independent of the cosmological models in the background.
Therefore, our work highlights the importance of considering the evolution of $\beta$ in the cosmology-fits.

In this paper, only the potential SN evolution is taken into account.
Some other factors, such as the evolution of $\sigma_{int}$ \cite{Kim2011},
may also cause systematic uncertainties for SNe Ia.
This issue deserves further study in future.

\begin{acknowledgments}
We are grateful to Dr. Alex Conley for providing us with the SNLS3 covariance matrices that
allow redshift-dependent $\beta$. We acknowledge the use of CosmoMC.
SW is supported by the Fundamental Research Funds for the Central Universities under Grant No. N130305007.
XZ is supported by the National Natural Science Foundation of China under Grant No. 11175042
and the Fundamental Research Funds for the Central Universities under Grant No. N120505003.
\end{acknowledgments}

\end{document}